\documentclass[aps,prd,showpacs,amssymb,superscriptaddress, notitlepage,floats,floatfix,nofootinbib, twocolumn,groupedaddress]{revtex4-1}

\usepackage{bm}
\usepackage{mathtools,amsmath,amssymb,amsfonts,eucal,mathrsfs,graphicx,tensor,csquotes,accents,commath,chngcntr}
\usepackage[dvipsnames]{xcolor}
\usepackage[unicode]{hyperref}
\hypersetup{colorlinks=true, citecolor=MidnightBlue,
            linkcolor=MidnightBlue, urlcolor=MidnightBlue, linktocpage=true}
\usepackage{journals}
\usepackage{wasysym}
\usepackage[normalem]{ulem}

\usepackage{fancyhdr}
\DeclareMathAlphabet{\mathpzc}{OT1}{pzc}{m}{it}
\definecolor{darkgreen}{rgb}{0.0, 0.6, 0.0}

\newcommand{\note}[1]{\text{\scshape\tiny{#1}}}
\newcommand{\mbf}[1]{\mathbf{#1}}
\newcommand{\gbf}[1]{\boldsymbol{#1}}
\newcommand{\ee}{\mathrm{e}}
\newcommand{\ii}{\mathrm{i}}
\newcommand{\dd}{\mathrm{d}}

\newcommand{\rh}{r_\mathrm{H}}

\newcommand{\al}{\alpha}
\newcommand{\be}{\beta}
\newcommand{\ga}{\gamma}

\newcommand{\de}{\delta}
\newcommand{\De}{\Delta}

\newcommand{\cep}{\varepsilon}
\newcommand{\ze}{\zeta}

\newcommand{\ka}{\kappa}
\newcommand{\la}{\lambda}
\newcommand{\La}{\Lambda}
\newcommand{\sg}{\sigma}

\newcommand{\om}{\omega}

\newcommand{\dl}{\partial}

\begin{document}

\title{Theory-agnostic Reconstruction of Potential and Couplings from Quasi-Normal Modes}

\author{Sebastian H. V\"olkel}
\email{svoelkel@sissa.it}
\author{Nicola Franchini}
\email{nfranchi@sissa.it}
\author{Enrico Barausse}
\email{barausse@sissa.it}
\affiliation{SISSA, Via Bonomea 265, 34136 Trieste, Italy and INFN Sezione di Trieste}
\affiliation{IFPU - Institute for Fundamental Physics of the Universe, Via Beirut 2, 34014 Trieste, Italy}

\date{\today}

\begin{abstract}

In this work, we use a parametrized theory-agnostic approach that connects the observation of black hole quasi-normal modes with the underlying perturbation equations, with the goal of reconstructing the potential and the coupling functions appearing in the latter. The fundamental quasi-normal mode frequency and its first two overtones are modeled through a second order expansion in the deviations from general relativity, which are assumed to be small but otherwise generic.
By using a principal component analysis, we demonstrate that percent-level measurements of the fundamental mode and its overtones can be used to constrain the effective potential of tensor perturbations and the coupling functions between tensor modes and ones of different helicity, without assuming an underlying theory. We also apply our theory-agnostic reconstruction framework to analyze simulated quasi-normal mode data produced within specific theories extending general relativity, such as Chern-Simons gravity.

\end{abstract}

\maketitle

\section{Introduction}\label{introduction}

After the first detection of the binary black hole (BH) merger GW150914~\cite{LIGOScientific:2016aoc}, the relentless experimental efforts conducted at the LIGO/Virgo 
interferometers have resulted in the detection of almost 100 more
compact binary mergers~\cite{LIGOScientific:2018mvr,LIGOScientific:2020ibl,LIGOScientific:2021djp}.
These detections will become more numerous as the sensitivity of these interferometers will increase and additional instruments will join the network (as recently done by KAGRA).
Besides a few exceptions involving neutron stars~\cite{LIGOScientific:2017ycc,LIGOScientific:2021qlt}, most of these events are binary BH mergers. On top of the astrophysical and cosmological implications that can be drawn from this growing experimental sample, there is also a significant interest in using it to test the validity of general relativity (GR) in the strong and dynamical regime~\cite{LIGOScientific:2016lio,LIGOScientific:2019fpa,LIGOScientific:2020tif,LIGOScientific:2021sio}. 

Many ongoing works aim to use the ringdown regime of binary BH mergers to conduct precision tests of the no-hair theorems' hypotheses (``BH spectroscopy''). By measuring multiple quasi-normal modes (QNMs),  quantitative experimental tests of the linearized perturbation equations of the Schwarzschild/Kerr space-time will become possible. Nevertheless, although perturbative calculations can be used to define the QNM spectrum as an eigenvalue problem, numerical relativity simulations are still needed to understand the range of validity of the perturbative regime, which is a problem still under development~\cite{Isi:2019aib,Giesler:2019uxc,Forteza:2020hbw,Cook:2020otn}.

In recent years, there have been several studies aiming to extract QNMs from real gravitational wave measurements~\cite{LIGOScientific:2016lio,LIGOScientific:2019fpa,LIGOScientific:2020tif,LIGOScientific:2021sio,Isi:2019aib,Giesler:2019uxc,Forteza:2020hbw,Cook:2020otn,Ghosh:2021mrv,Cotesta:2022pci}. 
The LIGO-Virgo collaboration reported that the $l=m=2$ and $n=0$ mode has been clearly extracted from GW150914~\cite{LIGOScientific:2016lio}, and follow up works~\cite{Isi:2019aib,Giesler:2019uxc} claim evidence also for the $n=1$ and $n=2$ overtones, although with much greater uncertainty  and a robustness still under debate~\cite{Cotesta:2022pci,Isi:2022mhy}. Under certain assumptions, it is  possible to combine measurements from different GW events~\cite{Ghosh:2021mrv}, providing more stringent bounds on possible deviations of the $l=m=2$ fundamental mode frequency from GR. The prospects to measure the $l=m=3$ mode have also been studied in Ref.~\cite{Cabero:2019zyt}.

Tests of GR, currently limited by the signal-to-noise ratio of the post-merger signal, will become easier with future, more sensitive detectors~\cite{Berti:2016lat}, although the increasing number of free parameters
needed to quantify deviations from GR
can be problematic~\cite{Bustillo:2020buq}. A task even more difficult than detecting a deviation from GR in the QNMs (if any) will be the extraction of information about the underlying theory or background metric. Although parametrized frameworks to capture modifications as function of BH mass and spin have been developed~\cite{Maselli:2019mjd,Carullo:2021dui}, they cannot be readily used to learn about the fundamental structure of the equations governing the perturbations,  nor about the underlying theory itself.

In general, a discrepancy with GR would manifest both in deviations
of the background BH metric from the Schwarzschild/Kerr solution~\cite{1916SPAW.......189S,Kerr:1963ud},
and in deviations from the linearized perturbation equations of GR
(i.e. the Regge-Wheeler/Zerilli equations for Schwarzschild~\cite{Regge:1957td,Zerilli:1970se} or the Teukolsky equation for Kerr~\cite{Teukolsky:1973ha}), see Ref.~\cite{Barausse:2008xv}.
Under the assumption that corrections to GR are ``small'', the deviations can be parametrized in the perturbation equations defining the QNM spectrum via a fixed set of theory-agnostic coefficients~\cite{Cardoso:2019mqo,McManus:2019ulj}.
This framework can then be used to attempt to solve an ``inverse problem'', {\it i.e.}~to determine the form of the perturbation equations from a given set of QNM observations.

In this work, we tackle this inverse problem in the non-rotating (spherical) case. 
More specifically, we focus on the possibility that the gravitational perturbations of different parity (axial and polar) may obey potentials deviating from GR and, moreover, that they can couple to a hypothetical scalar degree of freedom. While in GR a scalar degree of freedom is absent, it is a very common feature in many alternative theories of gravity~\cite{Berti:2015itd}, e.g. dynamical Chern-Simons gravity \cite{Alexander:2009tp} or degenerate higher-order scalar-tensor theories (DHOST) theories \cite{Langlois:2021aji}. 

In order to address the problem, we use and extend the parametrized framework introduced by Ref.~\cite{Cardoso:2019mqo,McManus:2019ulj} to handle QNM overtones, and produce a ``clean'' reconstruction by performing a principal component analysis (PCA)~\cite{Sivia2006,Pieroni:2020rob,Lara:2021zth}. The parametrized framework allows for a quick modeling of QNMs for small deviations from GR, while the PCA reveals the non-degenerate combinations of the parameters that can be extracted from the data.
We explicitly demonstrate the capabilities of our framework to constrain injected  deviations from GR in the effective potentials and coupling functions, provided that QNMs are known to within percent level.

This work is structured as follows. In Sec.~\ref{theoretical_minimum} we review BH perturbation theory and the parametrized framework of~\cite{Cardoso:2019mqo,McManus:2019ulj}; Sec.~\ref{methods} covers the details of the numerical and parameter estimation methods. The application and results are discussed in Sec.~\ref{application_results}. An overall conclusion is found in Sec.~\ref{conclusions}. Throughout this work, we use units in which $G=c=1$.

\section{Theoretical Framework}\label{theoretical_minimum}

The equations describing tensor and scalar perturbations of a non-spinning BH in GR are derived by linearizing respectively the Einstein and the Klein-Gordon equations, on top of a Schwarzschild geometry. These equations would depend in general on time, radius and angular coordinates. An expansion of the perturbation functions in spherical tensor or scalar harmonics eliminates the angular coordinates from the equations and decouples them. A solution to the resultant equation can be either found in the time domain, or, after performing a Fourier transform, in the frequency domain. The axial-parity equation for tensor perturbations is known as Regge-Wheeler equation~\cite{Regge:1957td}, whereas the Zerilli equation describes polar-parity tensor perturbations~\cite{Zerilli:1970se}. The perturbed Klein-Gordon equation for a scalar field takes a similar form. 
For useful reviews on the topic we refer the interested reader to Refs.~\cite{Kokkotas:1999bd,Nollert:1999ji,Berti:2009kk,Pani:2013pma}.

Working in the frequency domain,  the system of radial perturbation equations for $N_\note{f}$ coupled fields $\mbf{\Phi}$ of any helicity (tensor or scalar) around a spherically symmetric and static BH takes the general form
\begin{equation}\label{eq:mastersystem}
f \frac{\dd}{\dd r}\left( f \frac{\dd \mbf{\Phi}}{\dd r} \right) + \left[ \om^2 -f \mbf{V} \right]\mbf{\Phi} = 0.
\end{equation}
In this equation,
$r$ is the areal coordinate, $f=1-\rh/r$ (with 
$\rh$ the areal radius of the event horizon) 
 and $\om$ is the complex perturbation frequency. For each field, there is an infinite but discrete set of eigenfrequencies $\om_{n\ell}$, where $n$ is the overtone number 
 (characterizing the number of nodes of the radial solution) and $\ell$ its angular momentum number.
The diagonal terms of the matrix $\mbf{V}$ are the potentials felt by each field, while the non-diagonal ones represent coupling terms between fields.
We can thus write
\begin{align}
V_{ij} &= V^\note{GR}_{ij} + \delta V_{ij}, \\
\delta V_{ij} &= \frac{1}{\rh^2} \sum_{k=0}^{\infty} \alpha^{(k)}_{ij} \left(\frac{\rh}{r} \right)^k,\label{deltaVij}
\end{align}
where the $V^\note{GR}$ matrix is diagonal ($V^\note{GR}_{ii}\neq0$ and $V^\note{GR}_{ij}=0$ for $i\neq j$) and represents the GR potentials,
while the parameters $\alpha^{(k)}_{ij}$ are assumed to be small
and describe a generic deviation from GR.
In more detail, the GR potentials for scalar and tensor (axial and polar) modes are given by
\begin{align}
    V^\note{GR}_\note{scalar} & = \frac{\ell(\ell+1)}{r^2} + \frac{\rh}{r^3}\,, \\
    V^\note{GR}_\note{axial} & = \frac{\ell(\ell+1)}{r^2} - \frac{3\rh}{r^3}\,, \\
    V^\note{GR}_\note{polar} & = \frac{9\la \rh^2 r + 3 \la^2 \rh r^2 + \la^2(\la+2)r^3 + 9 \rh^3}{r^3 ( \la r + 3 \rh )^2 }\,,
\end{align}
where $\la = \ell (\ell+1) -2 $.

One can solve the eigenvalue problem for $\om$ described by Eq.~\eqref{eq:mastersystem}, with ingoing boundary conditions at the event horizon and outgoing ones at infinity. Different methods can be used for this purpose. In this paper,  in order to get the coefficients for the fundamental mode and the first two overtones, we employed a continued fraction method~\cite{Pani:2013pma,Leaver:1985ax,Rosa:2011my}, which provides more stable results than other methods, {\it e.g.,}~direct integration.~The details are explained in  Appendices~\ref{app:CF_1}-\ref{app:CF_2}.~Summarizing, the problem reduces to the computation of a complex function $\mathcal{L}\left(\om,\alpha^{(k)}_{ij}\right)$ for each choice of $i$, $j$ and $k$, and the corresponding eigenfrequencies $\om_{n\ell}$ correspond to the zeroes of this function. We denote the frequencies of the unperturbed GR problem, corresponding to $\alpha^{(k)}_{ij}=0$, as $\om^0_{n\ell}$, and we compute the non-GR frequencies through quadratic order in
the  $\alpha^{(k)}_{ij}$ coefficients as
\begin{equation}\label{eq:omega_mod}
    \om \approx \om^0 + \al^{(k)}_{ij} d_{(k)}^{ij} +  
    \frac{1}{2} \alpha^{(k)}_{ij} \alpha^{(s)}_{pq} e^{ijpq}_{(ks)} ,
\end{equation}
where we omitted the indices $n,\ell$ for readability.\footnote{In principle, there would be a quadratic contribution when the parameters $\alpha^{(k)}_{ij}$ depend on the frequency $\om$~\cite{McManus:2019ulj}. However, in our analysis we consider the parameters being independent from $\om$, therefore, we do not show this term here.}

One then needs to compute the coefficients $d_{(k)}^{ij}$ and $e^{ijpq}_{(ks)}$, which are independent from the specific deviation from GR under scrutiny, which is  encoded in the  parameters $\al^{(k)}_{ij}$ alone. In practice, one can determine  the coefficients by Taylor expanding the complex function $\mathcal{L}$ for small $\al^{(k)}_{ij}$. One can then show that the coefficients $d_{(k)}^{ij}$ and $e^{ijpq}_{(ks)}$ can be computed from a combination of the derivatives of the master function $\mathcal{L}$ with respect to the frequency and the parameters $\al^{(k)}_{ij}$, evaluated for $\al^{(k)}_{ij}=0$~\cite{McManus:2019ulj}.
We have checked the coefficients against those already computed in~\cite{Cardoso:2019mqo,McManus:2019ulj} and, despite the different method, they are in very good agreement.

\begin{table}
\[
\begin{array}{cccc}
\hline
 n & k & \rh d_{(k)} & \rh e_{(kk)} \\
\hline
 1 & 2 & 0.104137 - 0.004439 \ii & - 0.0149828 + 0.000895 \ii \\
   & 3 & 0.065239 - 0.010187 \ii & - 0.0048933 - 0.002945 \ii \\
   & 4 & 0.044246 - 0.000744 \ii & - 0.0032279 - 0.007499 \ii \\
   & 5 & 0.034315 + 0.008512 \ii & - 0.0017470 - 0.008550 \ii \\
   &10 & 0.014401 + 0.023307 \ii &   0.0069606 - 0.005194 \ii \\
\hline
 2 & 2 & 0.114665 + 0.000748 \ii & - 0.0202099 - 0.001664 \ii \\
   & 3 & 0.078288 - 0.013134 \ii & - 0.0138602 - 0.005958 \ii \\
   & 4 & 0.059947 + 0.001277 \ii & - 0.0163697 - 0.014995 \ii \\
   & 5 & 0.056594 + 0.016008 \ii & - 0.0163078 - 0.021431 \ii \\
   &10 & 0.048075 + 0.052281 \ii &   0.0139149 - 0.053472 \ii \\
\hline
\end{array}
\]
\caption{Linear and diagonal quadratic coefficients for axial gravitational perturbations, with different values of $k$ and $\ell = 2$.}
\label{tab:coeff}
\end{table}

\begin{figure}
\centering
\includegraphics[width=1\linewidth]{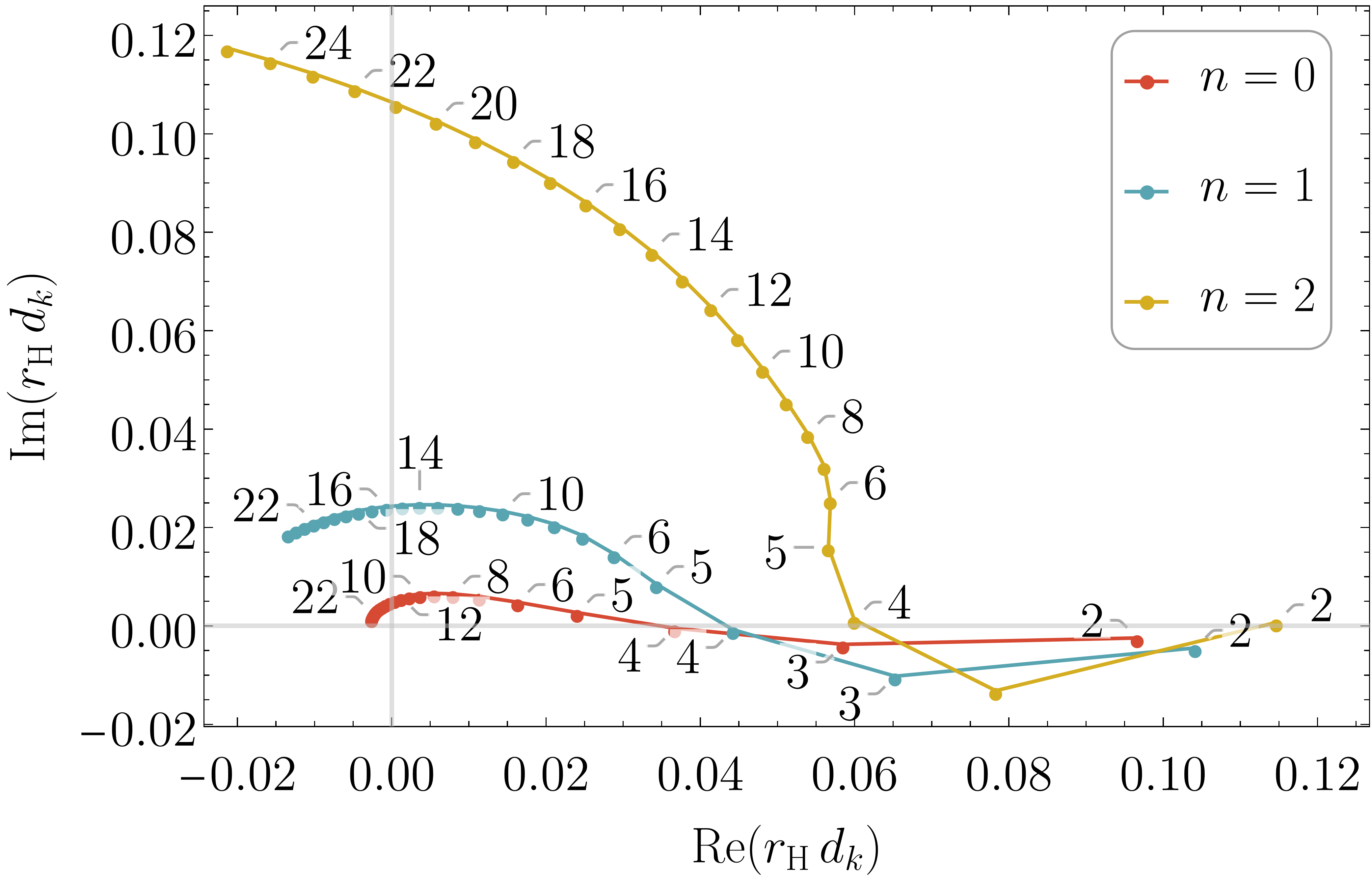}
\includegraphics[width=1\linewidth]{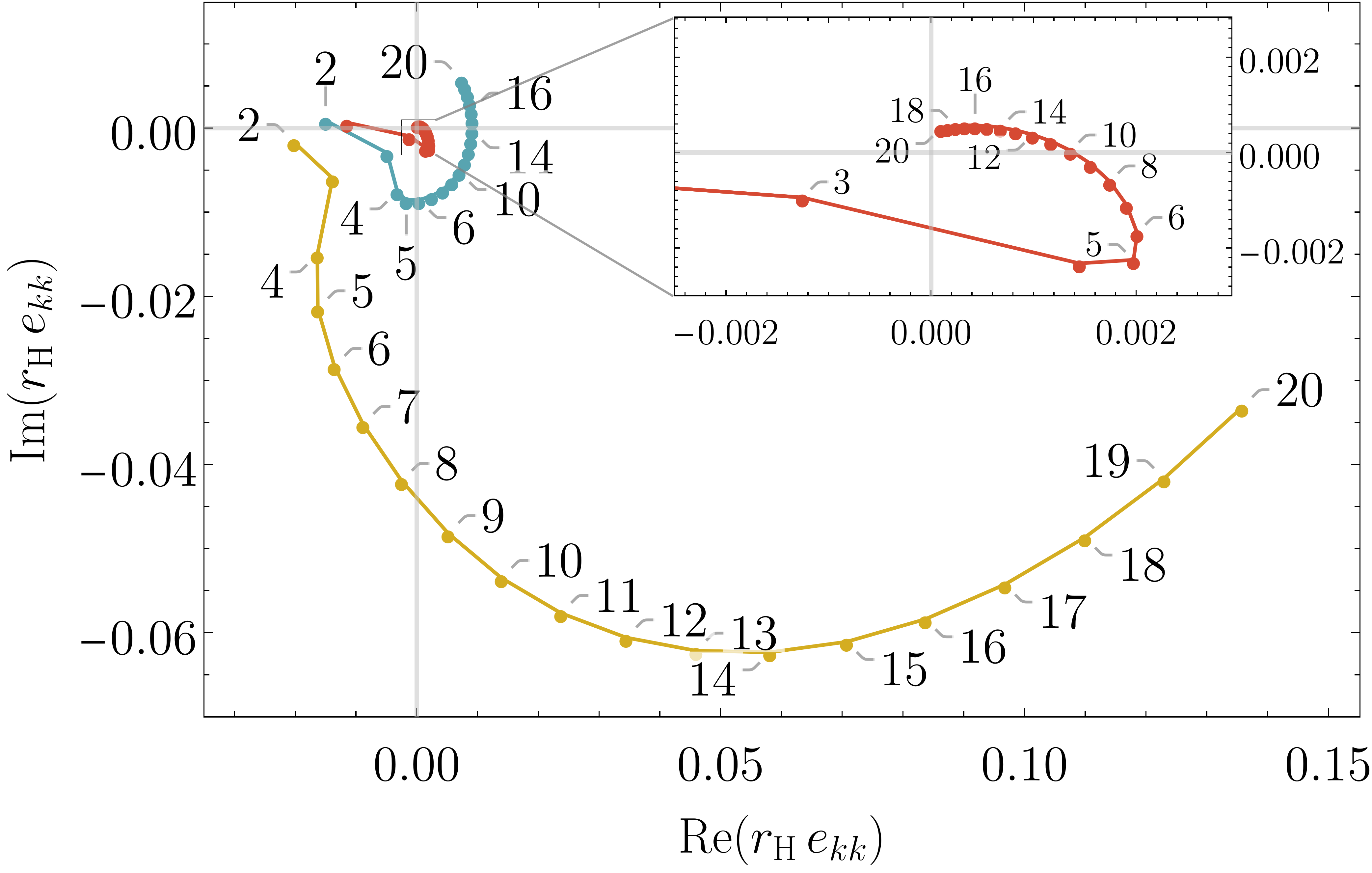}
\caption{Linear (top panel) and diagonal quadratic (bottom panel) coefficients for tensor axial QNMs and $\ell=2$. The upper panel assumes potential modifications in the indices $k \in [2,25]$, while $k \in [2,20]$ in the lower panel. Different overtones are represented with different colors ($n=0$: red, $n=1$: blue, $n=2$: yellow). \label{coeff_plot}
}
\end{figure}

We have extended the computation of the linear and quadratic coefficients, for both one field and two coupled fields, to the first two overtones $n=1,2$ for $\ell=2,3,4$. Table~\ref{tab:coeff} shows some of the coefficients for $\ell=2$ for axial tensor perturbations.
It is clear from the table that increasing the overtone number makes the coefficients grow for fixed $k$. To better visualize this behaviour, in Fig.~\ref{coeff_plot} we show the linear and diagonal quadratic coefficients, respectively $d_{(k)}$ and $e_{(kk)}$, for the axial case, plotting their real and imaginary part.
The case with  with $n=3$ is not shown in the plot, but follows the same trend.

\begin{figure}
\centering
\includegraphics[width=\linewidth]{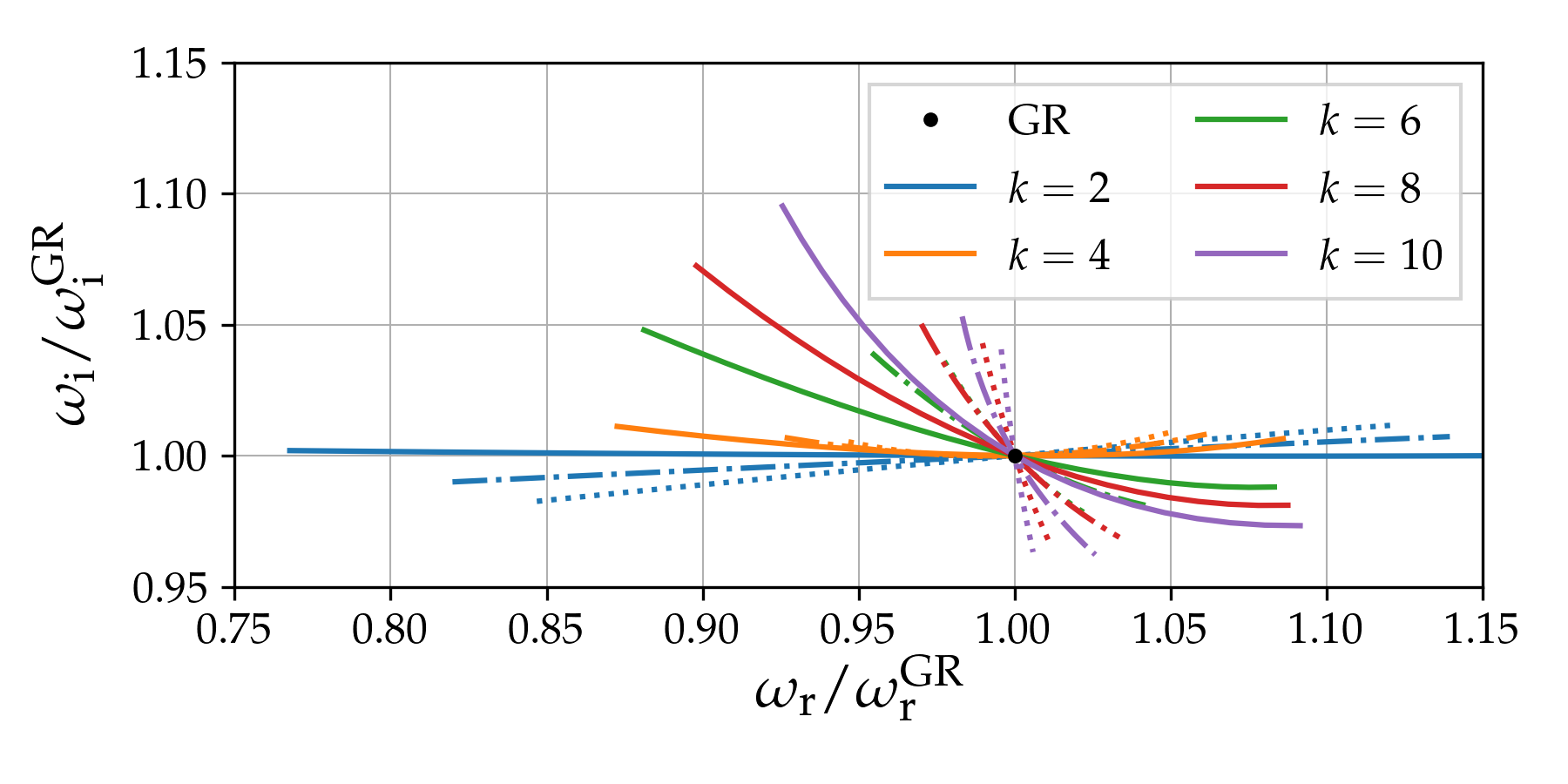}
\caption{Normalized QNM spectrum for the $l=2, n=0,1,2$ modes. Different colors correspond to different terms in the potential, while different styles correspond to different overtone numbers ($n=0$ dotted, $n=1$ dashed, $n=2$ solid). The minimum and maximum value for the magnitude of the potential modification was set to $\alpha^{(k)}=\pm1$. }
\label{qnm_art}
\end{figure}

This behavior directly affects the QNMs, as can be seen in Fig.~\ref{qnm_art}. Here, we choose a fixed value for $\al$ and compute the  modified frequencies with only one $k$ component (selected in the range $k\in[2,10]$). We display how the coefficients affect the real and imaginary $l=2$ frequencies, normalized by their GR values. Note that in GR the absolute value of the complex QNM frequencies increases with overtone number, which qualitatively implies that also the coefficients $d_{(k)}$ and $e_{(kk)}$ should increase, if they correspond to a roughly similar change in the QNM spectrum.

This behaviour suggests that higher modes are more sensitive to changes in the potential. We can provide a rough explanation of this with the Wentzel-Kramer-Brillouin (WKB)  approximation~\cite{Schutz:1985km,Iyer:1986np,Konoplya:2003ii}. This method connects the QNM frequencies to the derivatives of the effective potential with respect to the tortoise coordinate around its maximum $r_\mathrm{max}$. Studying the numerical precision of the WKB approximation implies that the higher the overtone, the 
more  derivatives of the effective potential one has to take into account.
 While modifications to the potential at its peak are proportional to $(\rh/r_\mathrm{max})^k=(2/3)^k$, its higher derivatives show a different decrease rate for large $k$, affecting  the magnitude of the overtones more significantly.

\section{Principal Component Analysis}\label{methods}

In this paper, we assume the observation of a certain number $N$ of QNMs, namely the fundamental mode $n=0$ and up to two more overtones $n=1,2$. 
We collectively denote them as the data $D_p$  and we
assume that they are measured with error $\sg^D_p$, with the index $p$ labeling
the different data points and running
 from $1$ to $2N$ (as we treat the real and imaginary parts of the QNM frequencies independently). 

The  model that we employ to describe the data is given by Eq.~\eqref{eq:omega_mod}. Each QNM frequency predicted by the model
is denoted by $F_p(\vec{\al})$, where $\vec{\al}$ is the vector containing all the parameters $\al^{(k)}_{ij}$.\footnote{To avoid cluttering the notation, we refer to the components of $\al$ as $\al_i$. The index $i$ runs from $1$ to $(k_\note{max}-k_\note{min})N_\note{f}$, being $k_\note{min}$ and $k_\note{max}$ the values determining the range of basis functions that we consider in Eq.~\eqref{deltaVij}.} Note that in general the parameters $\al^{(k)}_{ij}$ can be complex numbers, representing complex valued contributions from $\delta V_{ij}$. Although these can in principle arise for specific cases, for example when the potential becomes frequency dependent, we assume from here on that all $\al^{(k)}_{ij}$ are real numbers. A generalization to complex potentials is in principle straightforward, but introduces additional degeneracy to the inverse problem, which we will suppress implicitly in the following.

To construct the likelihood of the problem, we assume the QNM measurements to be uncorrelated and described by a Gaussian distribution with variance $\left(\sg^D_p\right)^2$. If the horizon location $\rh$ of the final BH were known, the likelihood would then be defined by a simple Gaussian distribution, whose logarithm 
would be proportional to
\begin{equation}\label{likelihood_nonmarg}
 \chi^2 = \sum_{p=1}^{2N} \left[ \frac{F_p(\vec{\al}) - D_p}{\sg^D_p} \right]^2.
\end{equation}
However, in general $\rh$ is not known, and it is therefore more robust to generalize Eq.~\eqref{likelihood_nonmarg} to take this into account.

In GR, $\rh=2 M$, and one could therefore try to estimate it from the measurement
of the individual BH masses during the inspiral~\cite{Barausse:2012qz}. However, 
the errors on the masses would propagate into the estimate of $\rh$. Moreover, beyond GR effects would generally make $\rh$ be different from $2M$. To model these uncertainties, we assume $\rh= \rh^{(0)}+\de \rh$, where 
$\rh^{(0)}$ is the GR expectation for $\rh$, and $\de \rh$ the deviation from it (due to errors in the measurement of the individual masses and 
to deviations from GR). By making the simplifying assumption that
$\de \rh$ is a Gaussian variable, we can write the likelihood as
\begin{equation}\label{likelihood_marg}
    P\left(D|\vec{\al},y\right) \propto \exp \left[ -\frac{\chi_y^2(D,\vec{\al})}{2} - \frac{y^2}{2\sg_y^2}\right],
\end{equation}
where
\begin{equation}
    \chi^2_y = \sum_{p=1}^{2N} \left[\frac{F_p(\vec{\al}) - (1+y)D_p}{\sg^D_p}\right]^2\,.
\end{equation}
Here, $y=\de \rh/\rh^{(0)}$
and $\sg_y$ is the relative error on $y$. The $1+y$ term multiplying $D_p$ 
arises from the scaling of 
 the observed frequencies
 with the 
BH horizon radius $\rh+\de\rh$.

To get rid of $y$ we can then marginalize over it, obtaining the  likelihood
\begin{equation}\label{marginalized_likelihood}
\begin{split}
    P(D|\vec{\al}) & = \int\dd y \, P\left(D|\vec{\al},y\right) \\
                   & = \frac{\sqrt{2\pi}}{B} \exp\left[ -\frac{\chi^2-A^2}{2} \right]\,,
\end{split}
\end{equation}
with
\begin{align}
    A & = \frac{1}{B} \sum_{p=1}^{2N} \frac{(F_p-D_p)D_p}{\left(\sg^D_p\right)^2}\,, \\
    B & = \sqrt{\frac{1}{\sg_y^2} + \sum_{p=1}^{2N} \left( \frac{D_p}{\sg^D_p} \right)^{2}}\,.
\end{align}
In the following, we assume $\sg_y = 5\,\%$. This choice is based on the underlying assumptions that the estimate on the final mass from the inspiral signal assuming GR will have small errors and that modifications to GR are small. This ensures that the location of the horizon will be approximated by $\rh\approx2M$.

With flat priors, the best-fit parameters (which we denote
by $\vec{\al}_0$) can be estimated from the maximum of the likelihood Eq.~\eqref{marginalized_likelihood}. We compute them by using the L-BFGS-B optimization method provided in the open source software package \textsc{SciPy} for \textsc{Python}~\cite{2020SciPy-NMeth}. We start from an initial guess for the parameters given by GR plus small random noise. To remain in the regime where the quadratic expansion  of Eq.~\eqref{eq:omega_mod} can be trusted, we bound the parameter search intervals for the parameters $\alpha_i$ to be of order unit.
Furthermore, by performing a quadratic expansion of the 
the log-likelihood near the maximum, one can obtain information on the errors
of the best-fit parameters. In more detail, the Hessian matrix
$\hat{H}$ evaluated at $\vec{\al}=\vec{\al}_0$,
\begin{align}
\left[ \hat{H}\right]_{ij}= -\frac{\dl^2}{\dl \al_i \dl \al_j}  \log \left[ P\left(D|\vec{\al}\right) \right],\label{hessian}
\end{align}
is the inverse of the covariance matrix of the parameters.
[Note the minus sign in Eq.~\eqref{hessian} to make the Hessian positive definite.]

In order to further clean the reconstruction of the potentials encoded 
in the best-fit parameters, we employ a technique called 
PCA~\cite{Sivia2006,Pieroni:2020rob,Lara:2021zth}, which we use to
``denoise'' the reconstruction of $\delta V_{ij}$.
The PCA allows one to find the linear combinations of the parameters $\vec{\al}$ that are best determined by a given set of QNMs.
This can be done by computing the eigenvalues $\lambda_k$ and eigenvectors $\hat{e}^\text{eig}_k$ of the Hessian Eq.~\eqref{hessian}. In this new basis, the eigenfunctions are orthogonal and their coefficients $b_k$ follow from projecting the best-fit parameters onto the eigenvectors:
\begin{align}
b_k = \vec{\al}_{0} \cdot \hat{e}^\text{eig}_k.
\end{align}
The errors on the coefficients $b_k$  are then given by the square root of the inverse of the eigenvalues, $\sigma_k = \lambda_k^{-1/2}$.

The ``denoising'' of the PCA  is then achieved by selecting only
the components that contribute significantly to the data. A possible criterion used {\it e.g.}~in Ref.~\cite{Pieroni:2020rob} consists of retaining only components for which
\begin{align}\label{PCA_criteria}
\frac{|b_k|}{\sigma_k} > i,
\end{align}
where $i=1$ allows only for eigenvectors that are not consistent with noise at 1-$\sigma$. This will select a set of $N^*$ eigenvectors.
The PCA reconstructed parameters $\vec{\al}_\text{PCA}$ are finally obtained as
\begin{align}
{\vec{\al}}_\text{PCA} = \sum_{k \in N^*} b_k \hat{e}^\text{eig}_k.
\end{align}
In appendix~\ref{disc3}, we will discuss  how different selection criteria affect the reconstruction.

With $\vec{\al}_\text{PCA}$ one can now compute the reconstruction of the potential as 
\begin{equation}
\de V_{ij}^\text{PCA} (r)= \de V_{ij}\left(r, \vec{\al}^\text{PCA}\right)\,,
\end{equation}
while the errors on the potential can be obtained by summing in quadrature
the errors on the retained coefficients $b_k$, which are Gaussian and uncorrelated 
\begin{align}\label{PCA_error}
\delta V_{ij}^\text{err}(r) = \sqrt{ \sum_{k\in N^*} \left[ \sigma_k \,\de V_{ij}\left(r, \hat{e}^\text{eig}_k\right) \right]^2}.
\end{align}

\section{Application and Results}\label{application_results}

In the following,  we consider various simulated QNMs as mock data, evaluated with full numerical calculations using the continued fraction method. As discussed in the Introduction, the $l=2$ and corresponding $n=0,1,2$ QNM frequencies are in principle within reach of current and future detectors. Hence, for the purposes of our reconstruction, a combined measurement of three frequencies simultaneously will be our optimistic assumption, while only one detected frequency our pessimistic assumption. 
For the sake of clarity, we have decided to focus in this section on results with constant relative errors, i.e. we assume that all  frequencies are measured within a $1\%$ error, unless stated otherwise. 
For better visualization of the injected and reconstructed potentials and coupling functions, we also introduce the compactified coordinate 
\begin{align}
x = 1-\frac{\rh}{r}.
\end{align}

\subsection{Reconstructing Potentials}\label{app1}

\begin{figure}
\centering
\includegraphics[width=1.0\columnwidth]{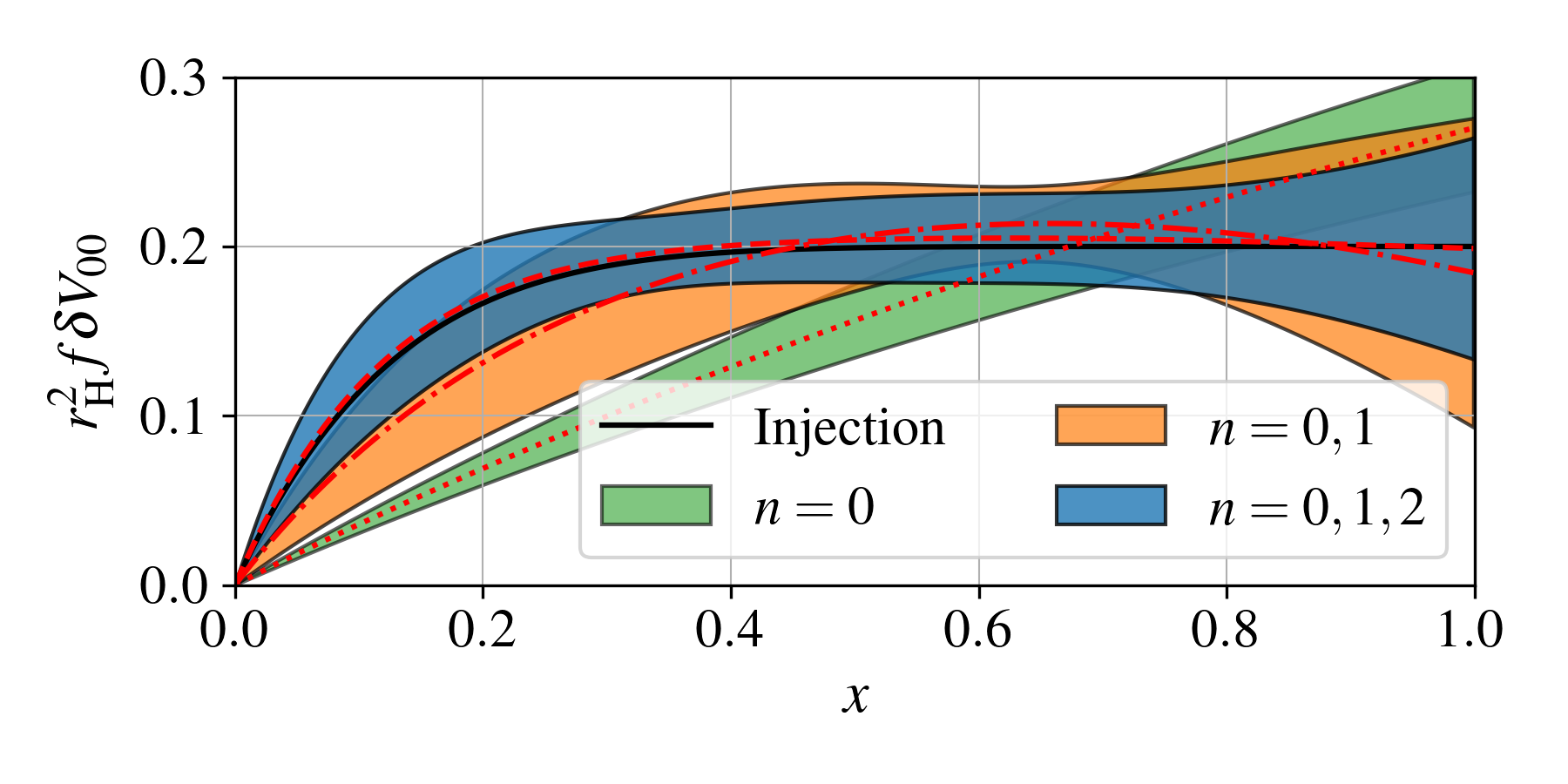}
\caption{Reconstruction of $\de V$ for an odd-parity tensor perturbation, with $\al^{(k)} = 0.2$ for $k\in [0,7]$, assuming measurements of $n=0,1,2$ modes with $1\%$ precision. The black solid line is the injection, while the red lines correspond to the PCA reconstruction using a different number of QNMs ($n=0$ dotted, $n=0,1$ dashed-dotted, $n=0,1,2$ dashed). The colored regions correspond to the PCA 2-$\sg$ errors of the reconstruction.
}
\label{Fig_app1_uncoupled}
\end{figure}

As a first example, we apply our framework to modifications in the axial potential only. The assumed injections are of the form $\al^{(k)}=0.2$ for $k \in [0,7]$, and we assume no couplings to additional fields. In the model given by Eq.~\eqref{deltaVij}, we also truncate the series at $k=7$.
The  2-$\sg$ PCA reconstruction region is shown in Fig.~\ref{Fig_app1_uncoupled}, where different colors correspond to different number of observed QNMs. 
Since GR corresponds to $\delta V_{00} = 0$, the injection can be clearly identified by the PCA and GR is excluded. This is already possible with 
knowledge of the fundamental mode alone, although the injection cannot be correctly identified in this case. It is evident that the inclusion of more QNMs significantly improves the reconstruction of the potential.
It is also worth noticing that although QNMs are very sensitive to the light ring region~\cite{Ferrari:1984zz,Schutz:1985km} (here around $x=1/3$), our PCA framework can clearly constrain the injection even far  from the BH, although with larger uncertainties.
Although not shown here, the results for  polar tensor fields and the scalar ones are quantitatively very similar.

\subsection{Reconstructing Coupling Functions}\label{app2}

\begin{figure}
\centering
\includegraphics[width=1.0\columnwidth]{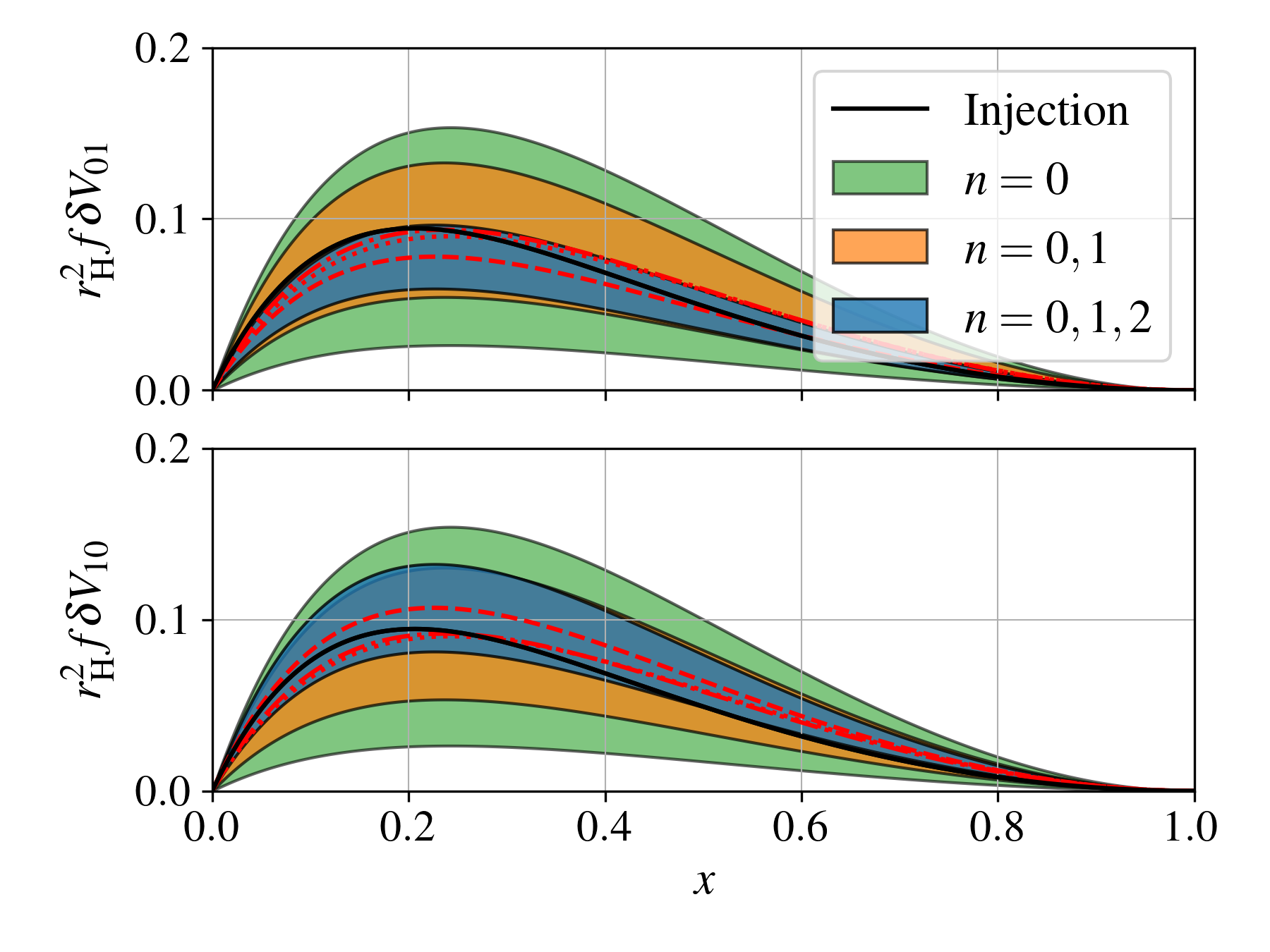}
\caption{Reconstruction of $\de V_{ij}$ for an axial gravitation perturbation coupled to a scalar field, with $\al^{(k)} = 0.2$ for $k\in [2,7]$, assuming measurement of $n=0,1,2$ modes with $1\%$ precision. The black solid line is the injection, while the red lines correspond to the PCA reconstruction using a different number of  QNMs ($n=0$ dotted, $n=0,1$ dashed-dotted, $n=0,1,2$ dashed). The colored regions correspond to the PCA 2-$\sg$ errors of the reconstruction. \label{Fig_app1_coupled}}
\end{figure}

As a next application, we consider the presence of a coupling function to a scalar field via $\delta V_{01}$ and $\delta V_{10}$. We assume injections only to originate from the two coupling functions, for which we assume $\al^{(k)}=0.2$ for $k \in [2,7]$. 
In principle, there can also be a contribution from $\delta V_{00}$, but we focus here on a case with purely GR potentials. For the PCA reconstruction, we let only the parameters of the coupling functions free to vary (also for $k \in [2,7]$).
The results for this case are shown in Fig.~\ref{Fig_app1_coupled}
and demonstrate that the fundamental mode alone is already enough to identify a significant non-GR contribution.

\begin{figure}
\centering
\includegraphics[width=1.0\columnwidth]{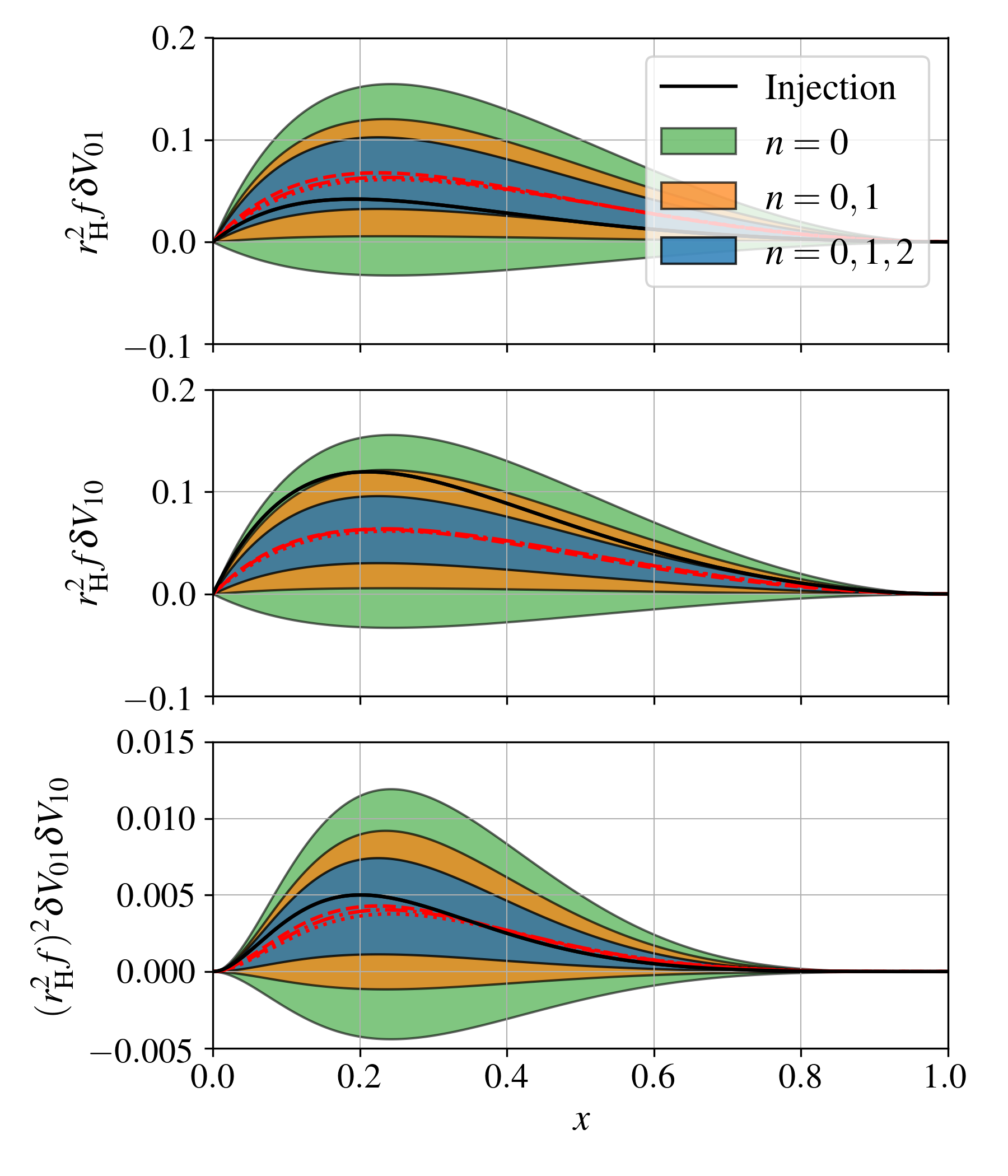}
\caption{Reconstruction of $\de V_{ij}$ for an axial gravitational perturbation coupled to a scalar field, with injection for $\alpha^{(k)}$ given in Table~\ref{tab:injection}, assuming measurement of $n=0,1,2$ modes with $1\%$ precision. The black solid line is the injection, while the red lines correspond to the PCA reconstruction using a different number of  QNMs ($n=0$ dotted, $n=0,1$ dashed-dotted, $n=0,1,2$ dashed). The colored regions correspond to the PCA 2-$\sg$ errors of the reconstruction. The top and middle panels show the individual coupling functions, while the bottom panel shows the reconstruction of their product.\label{Fig_app3_product_coupling}}
\end{figure}

The reconstruction of the coupling functions, however, can suffer from degeneracies when the injection is less trivial than in this first example. Let us then consider the injection given in Table~\ref{tab:injection}, where all the coefficients have different values. In 
the top and middle panels of Fig.~\ref{Fig_app3_product_coupling}, we show  the injected coupling functions and their reconstruction when both $\de V_{01}$ and $\de V_{10}$ are varied. As can be seen, the reconstruction -- obtained
by assuming the same free parameters in the model of Eq.~\eqref{deltaVij}
as in the previous example -- is suboptimal.
However, if one looks at the product of the coupling functions  (bottom panel)
the reconstruction improves significantly.
Moreover, it is worth noticing that in this example GR would be ruled out only with the measurement of three modes.

\begin{table}
\[
\begin{array}{c|cccccc}
    k & 2 & 3 & 4 & 5 & 6 & 7 \\
    \hline
    \al^{(k)}_{01} & 0.07 & 0.08 & 0.09 & 0.1 & 0.11 & 0.12  \\
    \al^{(k)}_{10} & 0.27 & 0.26 & 0.25 & 0.24 & 0.23 & 0.24 
\end{array}
\]
\caption{The injected parameters used in the application presented in Fig.~\ref{Fig_app3_product_coupling}.}
\label{tab:injection}
\end{table}

This degeneracy can be easily explained in two special limiting cases. The first is when both coupling functions have exactly one non-zero contribution $\bar{k}$, while all the others are zero. The second case is when all the contributions for the couplings have the same magnitude---{\it e.g.}~the one considered for Fig.~\ref{Fig_app1_coupled}. We denote these magnitudes by $ \al_{01}$ and $\al_{10}$. With a simple field redefinition, one can show that the QNMs in both cases are equivalent to those obtained by replacing $\al_{01} $ and $\al_{10} $ with the same coefficient $\al$, given by
\begin{align}\label{coupling_symmetry}
\al = \sqrt{ \al_{01} \al_{10}}.
\end{align}
From this result, one can infer that the reconstruction of the product of the coupling functions in these special cases must be invariant under the choice of injected parameters, for a fixed choice of $\al$. In the most general case, we found out that numerically evaluated spectra are still degenerate upon exchange $\de V_{01}$ and $\de V_{10}$. Even if we were not able to prove it formally, this fact strongly suggests that the product of the couplings is the best choice to reconstruct in any case.

\subsection{Dynamical Chern-Simons Gravity}\label{app4}

Finally, we consider the case of a specific theory: dynamical Chern-Simons (dCS) gravity~\cite{Alexander:2009tp}.  Here, the equations governing the perturbations around non-rotating BHs, which coincide with the Schwarzschild BH \cite{Yunes:2011we}, can be easily derived~\cite{Cardoso:2009pk,Molina:2010fb}. While the polar equation remains unchanged, the axial and scalar equations are coupled to each other, but the potentials are the same as in GR. As demonstrated in Ref.~\cite{McManus:2019ulj}, the axial QNMs are well approximated by
\begin{align}
\omega = \omega_0 + e^{1221}_{(55)} \left(12 \bar{\gamma} \sqrt{\pi \frac{(l+2)!}{(l-2)!}} \right)^2,
\end{align}
where  $\bar{\gamma}$ depends on the theory's coupling constant. Note that the latter can also be constrained by gravitational wave observations of the inspiral and mergers \cite{Stein:2013wza,Okounkova:2019dfo}.

\begin{figure}
\centering
\includegraphics[width=1.0\linewidth]{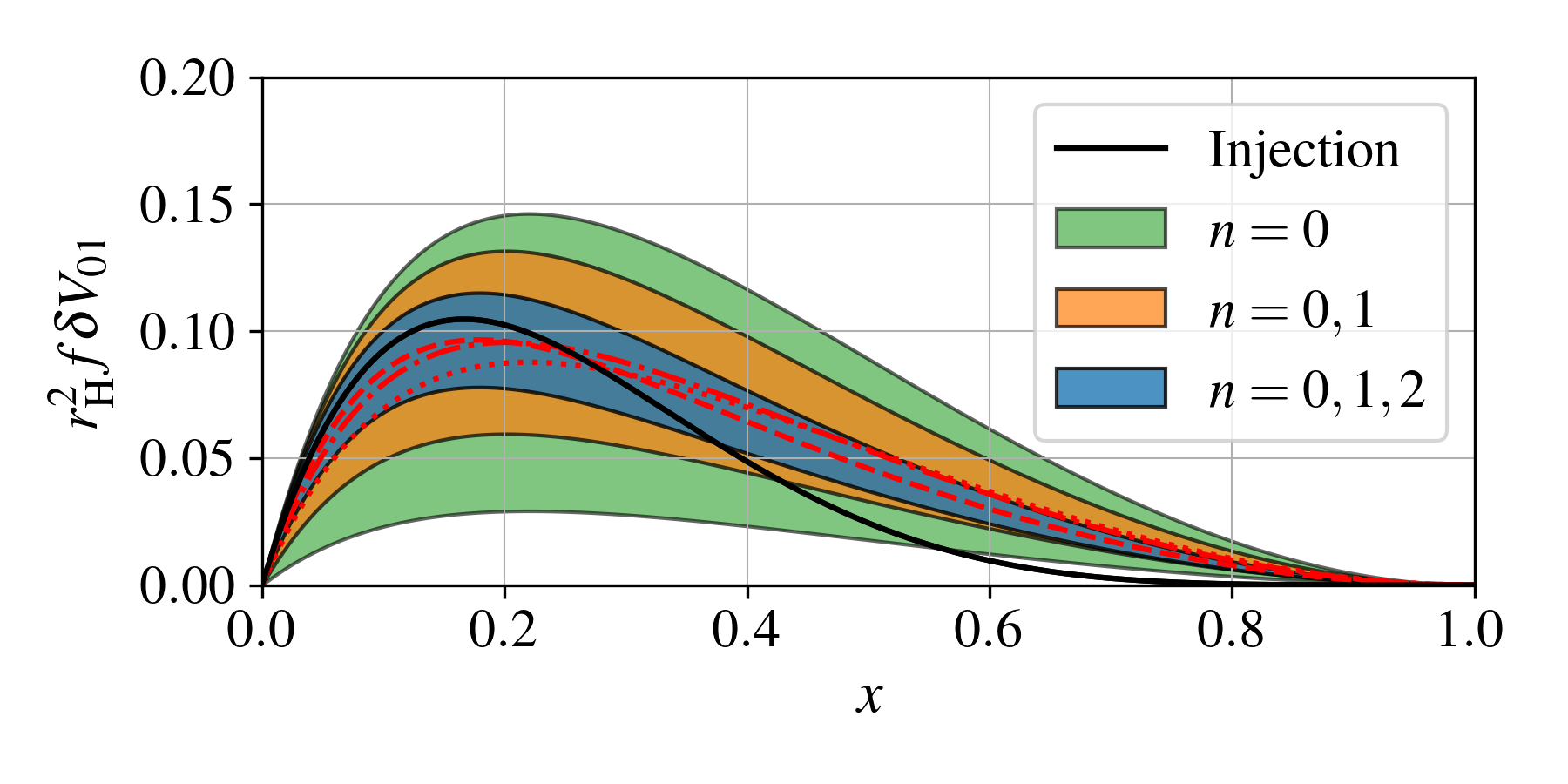}
\caption{PCA reconstructions of the dCS coupling function $\delta V_{01}$ assuming  QNM measurements with $1\%$ precision. The black solid line is the injection for $\bar{\gamma}=0.015$, while the red lines correspond to the PCA reconstruction using  different numbers of QNMs ($n=0$ dotted, $n=0,1$ dashed-dotted, $n=0,1,2$ dashed). The colored regions correspond to the PCA 2-$\sg$ errors of the reconstruction.}
\label{Fig_dcs_case}
\end{figure}

Since we already demonstrated the reconstruction of two independent coupling functions and the product of coupling functions, we now assume
that the coupling functions are symmetric, as indeed is the case for dCS. This time, we consider contributions up to $k=10$ in the model of Eq.~\eqref{deltaVij}. 
Our findings for $\bar{\gamma}=0.015$ are shown in Fig.~\ref{Fig_dcs_case}. Since we assume $\delta V_{01} = \delta V_{10}$ in this application, we only show $\delta V_{01}$. It can be clearly seen that our framework is capable of reconstructing the coupling functions and exclude GR. Moreover, the  reconstruction agrees very well with the injection in the region close to the BH, although the agreement degrades further out.

\subsection{Discussion}\label{discussion}

In all the cases considered above, we have focused on small modifications to the potentials and coupling functions of the scalar, axial and polar fields. As  pointed out in Ref.~\cite{McManus:2019ulj},  corrections to the QNM spectrum from the coupling functions enter at quadratic order in the deviations from GR, unless spectra are degenerate. Since quadratic corrections to the QNM spectrum also come from the individual potentials, it is in general very difficult to disentangle the two if they are allowed to vary at the same time. Indeed, we have tried to vary the potential and coupling functions at the same time, but the problem is very degenerate and sensitive to the initial guess of our root finder. This most likely happens as a result of possible multi-modalities in the likelihood/posteriors, calling for a more systematic approach (via {\it e.g.}~Markov Chain Monte Carlo or nested sampling).
We stress, however, that once the global maximum of the posteriors has been located, our PCA method can be applied in the vicinity of the maximum
to yield a denoised reconstruction.

\begin{figure}
\centering
\includegraphics[width=1.0\linewidth]{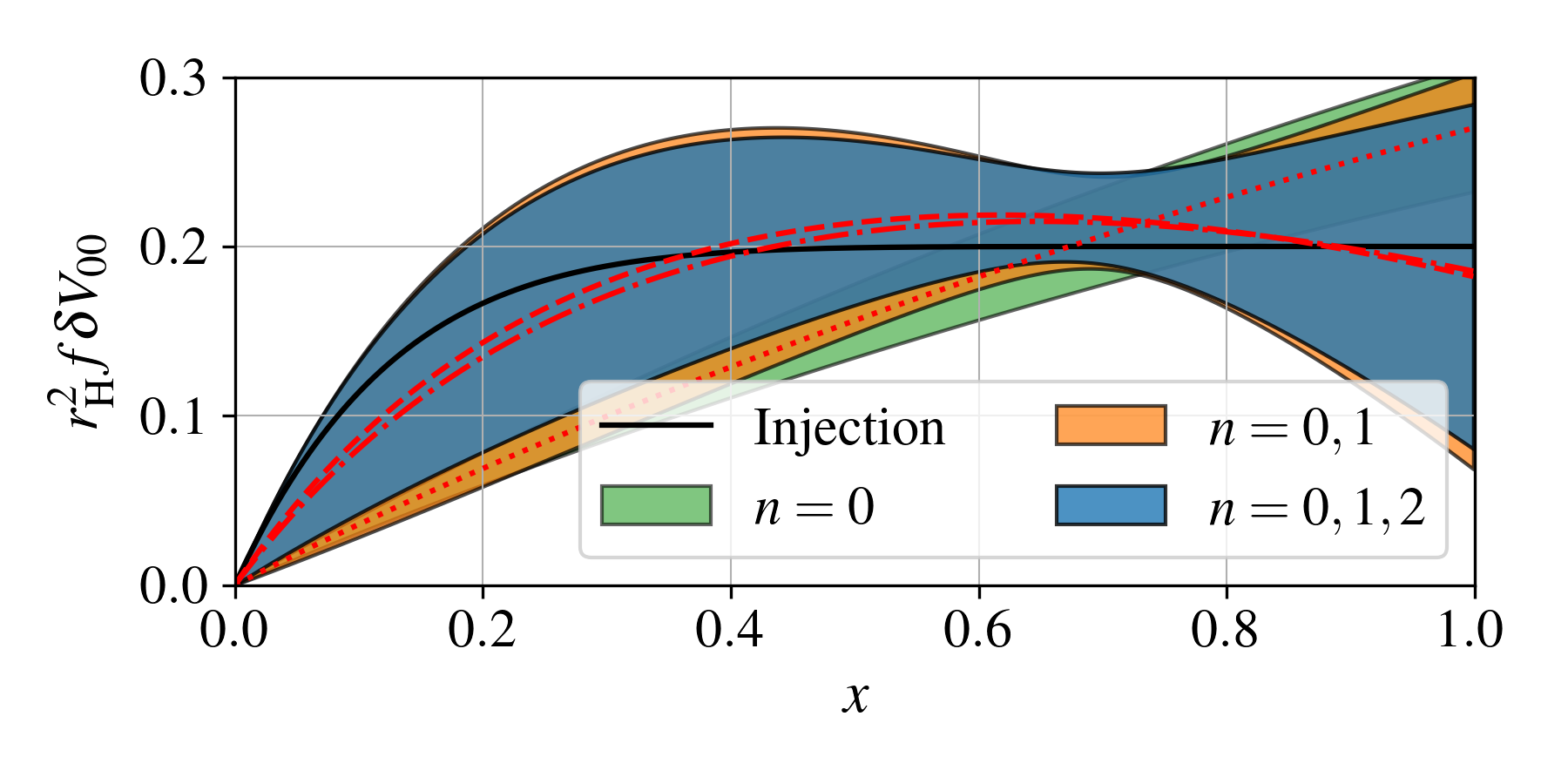}
\caption{PCA reconstruction for the same case as in Fig.~\ref{Fig_app1_uncoupled}, but with larger errors on the QNM overtones, i.e. of $1\,\%, 2\,\%, 5\,\%$ for the different overtones ($n=0,1,2$), respectively. The black solid line is the injection, while the red lines correspond to the PCA reconstruction using a different number of  QNMs ($n=0$ dotted, $n=0,1$ dashed-dotted, $n=0,1,2$ dashed). The colored regions correspond to the PCA 2-$\sg$ errors of the reconstruction.}
\label{PCA_unevenerrors}
\end{figure}

Another point that can have an impact on the reconstruction of the potentials is the precision with which the QNMs are known. In the analysis conducted so far, we showed how the reconstruction of the potentials improves when multiple QNMs are provided. However, in that application we assumed optimistic $1\%$ errors on  measured frequencies.
We now compare the reconstruction of the potential in section~\ref{app1} with
one that assumes larger uncertainties for the overtones. In more detail, we now assume that the $n=0,1,2$ QNMs are known with $1\%, 2\%, 5\%$ relative errors, respectively, while the other assumptions are unchanged. The resulting reconstruction is shown in Fig.~\ref{PCA_unevenerrors}. By comparing with Fig.~\ref{Fig_app1_uncoupled}, it is clear that the reconstruction still approximates well the injection, although the improvement allowed by including the second overtone is now  marginal.
Nevertheless, one can still confirm a non-GR modification of the potential at high significance.

One may ask how the picture changes if instead of providing multiple overtones $n$ for the same $l$, one provides the $n=0$ modes for multiple $l$, see e.g. \cite{Gossan:2011ha}. In the current parametrization framework each potential is treated independently because each set of deviations $\alpha_k$ implicitly depends on $l$. This implies one cannot easily combine the QNMs to obtain a better reconstruction. An alternative approach could be to provide a parametrized BH metric and assume a certain structure of the perturbation equations, as done in Ref.~\cite{Volkel:2020daa}. Here one could combine QNMs of different $l$ to constrain the same (metric) parameters.

Finally, the reconstruction depends mildly also on the PCA criteria used to 
select what components are retained in the reconstruction. We discuss this in detail in Appendix~\ref{disc3}.

\section{Conclusions}\label{conclusions}

In this work, we have tackled the `inverse problem' of reconstructing
the potential of the gravitational perturbations on a spherically symmetric static background from a set of simulated
observations of QNMs. We consider both the case of tensorial perturbations
only, and that of coupled tensorial and scalar gravitons. In the latter case,
we attempt to reconstruct also the coupling functions between the two helicities.
Our approach parametrizes the potential and coupling functions in a theory independent way introduced in Refs.~\cite{Cardoso:2019mqo,McManus:2019ulj}, which we 
generalized to the $n=1,2$ overtones. In this approach, the only assumption is that the deviations of the gravitational theory from GR are small, and the QNM frequencies are computed through quadratic order in these deviations. 
Since the problem of reconstructing the potentials and coupling functions 
in a theory independent formalism is intrinsically degenerate, we employ a PCA to
`denoise' the reconstruction. The PCA technique achieves by identifying the `modes' or `features' of the potentials and coupling functions that are best constrained by the data.

Unlike the perturbation potential, which modifies the QNMs already at linear order in the deviations from GR, the presence of couplings between scalar and tensor modes affects the QNM spectrum only at quadratic order. This makes it more difficult to reconstruct coupling functions. Nevertheless, we have shown that given QNM frequency measurements at percent level or better, one can successfully reconstruct at least suitable combinations (products) of the coupling functions.
In general, in most of the cases that we considered, our PCA framework allows for the extraction of two to three features, especially if QNM overtones are measured. 

In an earlier work, some of us performed a similar analysis of the inverse problem using Bayesian techniques and higher order WKB theory for the computation of QNMs~\cite{Volkel:2020daa}.
The latter allowed us to go beyond small modifications from GR, but did not include couplings between modes of different helicity.
That work also managed to account for non-trivial correlations between possible modifications of the background space-time and the perturbation equations. We plan to perform a similar reconstruction of the space-time metric (but allowing also the potentials and coupling functions to vary) in future extensions of this paper.
This would also open the way to combining information on the background BH geometry
from QNM observations with related information from
electromagnetic probes, such as X-rays spectra~\cite{Bambi:2016sac,Tripathi:2018lhx} and 
shadow measurements by the Event Horizon Telescope~\cite{EventHorizonTelescope:2019dse,EventHorizonTelescope:2019ggy}. Indeed, the latter have been the object of several recent studies to constrain possible deviations of the BH geometry from GR, {\it e.g.}~\cite{EventHorizonTelescope:2020qrl,Volkel:2020xlc,EventHorizonTelescope:2021dqv,Nampalliwar:2021oqr,Kocherlakota:2022jnz,Lara:2021zth}. 

Future developments  of our formalism will also include the use of real ringdown data, as well as an extension to the case of rotating BHs. The latter is particularly challenging because even theory specific calculations of QNMs in rotating background are currently limited to the slow rotation approximation~\cite{Cano:2020cao,Pierini:2021jxd,Wagle:2021tam}.
Moreover, a different parametrization of the potentials and coupling functions
may allow for describing near-horizon deviations from GR~\cite{Cardoso:2016rao,Cardoso:2017cqb}, or the presence of
matter sources far from the BH~\cite{Barausse:2014tra,Cheung:2021bol}.
While the deviations produced in the QNM spectrum can be sizeable in these cases, the time-domain ringdown signal is less significantly affected~\cite{Barausse:2014tra,Nollert:1996rf,Nollert:1998ys,Jaramillo:2020tuu}. Therefore, our formalism could be applicable in the time domain.

\acknowledgments
We would like to thank Emanuele Berti, Andrea Maselli, Ryan McManus and Guillermo Lara for useful discussions.
We acknowledge financial support provided under the European Union's H2020 ERC Consolidator Grant ``GRavity from Astrophysical to Microscopic Scales'' grant agreement no. GRAMS-815673. This work was supported by the EU Horizon 2020 Research and Innovation Programme under the Marie Sklodowska-Curie Grant Agreement No. 101007855.

\bibliography{literature}

\appendix

\section{Continued fraction method - Single field}\label{app:CF_1}

In this section we revisit the continued fraction method applied to the problem of our interest. In order to solve Eq.~\eqref{eq:mastersystem} for a single field $\Phi$, one needs to assume the following expansion for the field
\begin{equation}\label{eq:CF_ansatz}
    {\Phi} = \frac{\ee^{-r \ka}}{ r^{\chi}} f^{\rho} \sum_{n=0}^N {a}_{n} f^n\,,
\end{equation}
where $N$ is an arbitrary large number, and, in order to impose proper boundary conditions, we assume $\ka = \rho\zeta $ and $\chi = \left(\al^{(0)}+\al^{(1)} + 2\rho^2\right)/2\ka$, where $\rho=-\ii\om$ and $\ze=\sqrt{1+{\al^{(0)}}/{\rho^2}}$. Substituting this ansatz into the equation of motion, one gets the following relation between the coefficients ${a}_n$
\begin{equation}
    \sum_{m=-1}^q {A}^{(m)}_{n} {a}_{n-m} = 0,
\end{equation}
where the form of the coefficients ${A}^{(m)}_{n}$ depends on the specific problem, as well as $q$. For scalar perturbation and tensor axial perturbation $q = \max \left( K-2, 2 \right)$, while for tensor polar perturbations $q = \max \left( K, 3 \right)$. In both cases, $K$ is the integer at which we truncate the expansion in $1/r$ powers inside $\de V$. For the scalar and axial case, the coefficients will take the form
\begin{align}
    A^{(-1)}_n & = \al_n \,, \\
    A^{(0)}_n & = \be_n - H + \sum_{k=2}^K\De^{(0)}_{(k)}  \,, \\
    A^{(1)}_n & = \ga_n + H + \sum_{k=2}^K\De^{(1)}_{(k)}  \,, \\
    A^{(m)}_n & = \sum_{k=2}^K\De^{(m)}_{(k)}\,, \qquad \qquad \text{with } 2\leq m\leq q.
\end{align}
The coefficients read explicitly $\al_n = n(n+2\rho)$, $\De^{(m)}_{(k)} = \al^{(k)}\left(-1\right)^{m+1} \binom{k-2}{m}$, $H = 1$ for scalar perturbation and $H = -3$ for tensor axial perturbation and
\begin{widetext}
\begin{align}
\notag
    \be_n = & -\zeta ^2 \rho ^2-\Lambda -2 n^2+{\al^{(1)}} \left(\frac{-2 n-2 \rho +1}{2
   \zeta  \rho }-1\right)-\frac{3}{2} \zeta  \rho  (2 n+2 \rho -1) \\
   & \label{eq:CF_beta} -\frac{\rho 
   (2 n+2 \rho -1)}{2 \zeta }+n (2-4 \rho )-3 \rho ^2+2 \rho  \,, \\
   \label{eq:CF_gamma}
   \ga_n = & \left[\frac{{\al^{(1)}}+(\zeta +1)^2 \rho^2 +2 \zeta \rho  (n-1)}{2 \zeta  \rho } \right]^2 -1  \,.
\end{align}

On the other hand, the structure for tensor polar perturbations is
\begin{align}
    A^{(-1)}_n & = \al_n  \,, \\
    A^{(0)}_n  & = \be_n + \sum_{k=2}^K\De^{(0)}_{(k)}  \,, \\
    A^{(1)}_n  & = \ga_n + \sum_{k=2}^K\De^{(1)}_{(k)}  \,, \\
    A^{(2)}_n  & = \de_n + \sum_{k=2}^K\De^{(2)}_{(k)}  \,, \\
    A^{(3)}_n  & = \cep_n + \sum_{k=2}^K\De^{(3)}_{(k)}  \,, \\
    A^{(m)}_n  & = \sum_{k=2}^K\De^{(m)}_{(k)}\,, \qquad \qquad \text{with } 4\leq m\leq q  \,.
\end{align}

The coefficients are $\al_n = n(n+2\rho)$ and

\begin{align}
\notag
    \be_n = & -\zeta ^2 \rho ^2-\frac{\Lambda ^2+\Lambda  \left(3 \rho ^2-2 \rho -2\right)+2 (\Lambda +4) n^2+2 n \left[\Lambda  (2 \rho -1)+8 \rho -7\right]+3 \rho ^2-14 \rho +9}{\Lambda
   +1} \\
   & + \al^{(1)}\frac{ -2 (\zeta +1) \rho -2 n+1}{2 \zeta  \rho }-\frac{3}{2} \zeta  \rho  (2 n+2 \rho -1)-\frac{\rho  (2 n+2 \rho -1)}{2 \zeta }\,, \\
   \notag
   \ga_n = & \frac{{\al^{(1)}}^2}{4 \zeta ^2 \rho ^2}  + \al^{(1)} \left[\frac{(\zeta +1) \left[\zeta  (\Lambda +13)+\Lambda +1\right]}{2 \zeta ^2
   (\Lambda +1)}-\frac{\Lambda -\Lambda  n-7 n+10}{\rho \zeta (  \Lambda +1 )}\right]\,, \\
   \notag
   & +\frac{(\zeta +1)^3 \rho ^2 \left[\zeta  (\Lambda +25)+\Lambda +1\right]}{4 \zeta ^2 (\Lambda +1)}+\frac{\left[\Lambda  (\Lambda +14)+22\right] n^2-2 \left[\Lambda  (\Lambda +20)+37\right] n + 3 (\Lambda +3) (\Lambda
   +7)}{(\Lambda +1)^2} \\
   & +\rho  \left[\frac{3 (3 \zeta +1) (\zeta +1) (2 n-3)}{\zeta  (\Lambda
   +1)}+\frac{(\zeta +1)^2 (n-1)}{\zeta }+\frac{18 (n-2)}{(\Lambda +1)^2}\right]\,, \\
   \notag
   \de_n = & -\frac{3 {\al^{(1)}}^2}{2 \zeta ^2 (\Lambda +1) \rho ^2}+\al^{(1)}
   \left[\frac{24 \Lambda -6 (2 \Lambda +5) n+69}{2 \zeta  (\Lambda +1)^2 \rho }-\frac{3 (\zeta +1) \left[\zeta  (\Lambda +4)+\Lambda +1\right]}{\zeta ^2 (\Lambda +1)^2}\right] \\
   \notag
   & -\frac{3 (\zeta +1)^3 \rho ^2 \left[\zeta  (\Lambda +7)+\Lambda +1\right]}{2 \zeta ^2 (\Lambda +1)^2} +\frac{3
   (\zeta +1) \rho  \left[8 (\zeta +1) \Lambda +53 \zeta -2 n (2 (\zeta +1) \Lambda +11 \zeta +5)+23\right]}{2 \zeta  (\Lambda +1)^2} \\
   & -\frac{3 \left[9 (\Lambda +5)+2 n (-4 \Lambda +(\Lambda
   +4) n-19)\right]}{(\Lambda +1)^2}\,, \\
   \cep_n  = & \frac{9 \left[\al^{(1)}+\rho  \left[\zeta ^2 \rho +2 \zeta  (n+\rho -3)+\rho \right]\right]^2}{4 \zeta ^2 (\Lambda +1)^2 \rho ^2}\,, \\
   \De^{(m)}_{(k)} = & \al^{(k)}(-1)^{m+1} \left[ \binom{k-2}{m} + \frac{6}{\La+1}\binom{k-2}{m-1} + \frac{9}{(\La+1)^2} \binom{k-2}{m-2} \right] \,.
\end{align}

When $q\geq 2$, one can perform $q-1$ steps of Gaussian elimination to get a three term recurrence relation between the coefficients $a_n$. The relation between the coefficients at the $p$-th elimination step is
\begin{equation}
    A_{n,p}^{(m)} = 
    \begin{cases}
    A_{n,p - 1}^{(m)} & \text{if  $n < q-p+1$, or $m > q-p+1$}\\
    A_{n,p - 1}^{(m)} - \frac{A^{(q-p+1)}_{n,p-1}}{A^{(q-p)}_{n-1,p}}A_{n-1,p}^{(m-1)} & \text{else.}
    \end{cases}
\end{equation}
\end{widetext}
The final three-terms relation reads as
\begin{align}
    &\al_0 a_1 + \tilde{\be}_0 a_0 = 0 \, \\
    & \al_n a_{n+1} + \tilde{\be}_n a_n + \tilde{\ga}_n a_{n-1} = 0 \quad \text{if $n\geq 1$,}
\end{align}
where $\tilde{\be}_n = A^{(0)}_{n,q-1}$ and $\tilde{\ga}_n = A^{(1)}_{n,q-1}$.
The Leaver method works as follows: one can construct the $n$-th ladder operator $R_n$ from the next one as
\begin{equation}
 R_n = \frac{\tilde{\ga}_n}{\tilde{\be}_n - \al_n R_{n+1}}\,,
\end{equation}
where the operator has the property $a_{n+1} = R_n a_n$. By initializing arbitrarily $R_n$ for a large value of $n$, one can compute the step to find the equation
\begin{equation}
    \mathcal{L}\left(\om, \al^{(k)}\right) \equiv R_1 - \frac{\tilde{\be}_0}{\al_0} = 0 \,.
\end{equation}
The roots of this equation are the eigenfrequencies of the problem.

\section{Continued fraction method - Two or more fields}\label{app:CF_2}

In this section of the appendix, we sketch the idea behind the continued fraction method applied to the problem~\eqref{eq:mastersystem} with two fields. It was inspired by the multi-field application of this method exposed in~\cite{Pani:2013pma,Rosa:2011my}. For simplicity we assume a coupling between scalar propagation and tensor axial propagation. The procedure can be straightforwardly generalized to more fields, and different helicities.

For both the axial and the scalar field, we assume an ansatz of the form of Eq.~\eqref{eq:CF_ansatz}, though, taking different choice of the coefficients $a_n$ for the two fields, namely, $a_n^\note{tensor}$ and $a_n^\note{scalar}$. We store these coefficients into the two-dimensional vectors $\mbf{U}_n = \left( a_n^\note{tensor}, a_n^\note{scalar} \right)$. Hence, from the system of equations, we can infer the following relation between coefficients
\begin{equation}
    \sum_{m=-1}^q \mbf{A}^{(m)}_n \mbf{U}_{n-m} = 0,
\end{equation}
where $q= \max\left( K-2,2 \right)$, and the matrices $\mbf{A}^{(m)}_n$ read
\begin{align}
    \mbf{A}^{(-1)}_n & = 
    \begin{pmatrix}
    \al_n & 0 \\
    0 & \al_n
    \end{pmatrix} \,, \\
    \mbf{A}^{(0)}_n & = \sum_{k=2}^K
    \begin{pmatrix}
    \be_n + 3 + \De^{(0)}_{(11k)} & \De^{(0)}_{(12k)} \\
    \De^{(0)}_{(21k)} & \be_n - 1 + \De^{(0)}_{(22k)}
    \end{pmatrix} \,, \\
    \mbf{A}^{(1)}_n & = \sum_{k=2}^K
    \begin{pmatrix}
    \ga_n - 3 + \De^{(1)}_{(11k)} & \De^{(1)}_{(12k)} \\
    \De^{(1)}_{(21k)} & \ga_n + 1 + \De^{(1)}_{(22k)}
    \end{pmatrix} \,, \\
    \mbf{A}^{(m)}_n & = \sum_{k=2}^K
    \begin{pmatrix}
    \De^{(m)}_{(11k)} & \De^{(m)}_{(12k)} \\
    \De^{(m)}_{(21k)} & \De^{(m)}_{(22k)}
    \end{pmatrix} \,,
\end{align}
where $\al_n = n(n+2\rho)$, $\be_n$ and $\ga_n$ are the same coefficients of Eq.~\eqref{eq:CF_beta}-\eqref{eq:CF_gamma}, and $\De^{(m)}_{(ijk)} = \al^{(k)}_{ij}\left(-1\right)^{m+1} \binom{k-2}{m}$. Due to the ordering that we chose, the index $i=j=1$ in $\al^{(k)}_{ij}$ refers to tensor field, and $2$ to scalar field.

The Gaussian elimination works as for the single field case. In order to get a three-terms relation, one can perform $q-1$ steps of elimination. The $p$-th step reads as
\begin{widetext}
\begin{equation}
\mbf{A}_{n,p}^{(m)} = 
\begin{cases}
    \mbf{A}_{n,p - 1}^{(m)} -\mbf{A}^{(n-p+1)}_{n,p-1} \left( \sum_{j=p}^{n-1}\mbf{A}^{(j-p+1)}_{j,p-1}\right)^{-1} \left(\sum_{j=p}^{n-1}\mbf{A}^{(j-n+m)}_{j,p-1}\right) & \text{for $2\leq n-p+1\leq q$,} \\
    \mbf{A}_{n,p - 1}^{(m)}  & \text{else.} \\
\end{cases}
\end{equation}
\end{widetext}
The final three-terms relation reads as
\begin{align}
    &\tilde{\gbf{\al}}_0 \mbf{U}_1 + \tilde{\gbf{\be}}_0 \mbf{U}_0 = 0 \, \\
    &\tilde{\gbf{\al}}_n \mbf{U}_{n+1} + \tilde{\gbf{\be}}_n \mbf{U}_n + \tilde{\gbf{\ga}}_n \mbf{U}_{n-1} = 0  \quad \text{if $n\geq 1$,}
\end{align}
where $\tilde{\gbf{\al}}_n = \mbf{A}^{(-1)}_{n}$, $\tilde{\gbf{\be}}_n = \mbf{A}^{(0)}_{n,n-1}$ and $\tilde{\gbf{\ga}}_n = \mbf{A}^{(1)}_{n,n-1}$.
Analogously to the single-field case, one can construct the $n$-th ladder operator $\mbf{R}_n$ from the next one as
\begin{equation}
 \mbf{R}_n = \tilde{\gbf{\ga}}_n \left(\tilde{\gbf{\be}}_n - \gbf{\al}_n \mbf{R}_{n+1}\right)^{-1}\,,
\end{equation}
where, again, the operator has the property $\mbf{U}_{n+1} = \mbf{R}_n \mbf{U}_n$. The final equation whose roots are the eigenfrequencies of the problem is
\begin{equation}
    \mathcal{L}\left(\om, \al^{(k)}_{ij}\right) \equiv \det \left[\mbf{R}_1 - \tilde{\gbf{\be}}_0 \times \left(\gbf{\al}_0\right)^{-1} \right]= 0 \,.
\end{equation}

\section{Relevance of PCA Criteria}\label{disc3}

The two main aspects that affect the shape of the PCA reconstruction are the number of overtones observed and the criterion used to select the relevant components. 
Throughout the reconstruction analysis of section~\ref{application_results}, we made use of the PCA criteria defined in Eq.~\eqref{PCA_criteria}. Instead, in this section only we fix the number of relevant modes, and compare the reconstructions for different number of observed modes.

\begin{figure}
\centering
\includegraphics[width=1.0\linewidth]{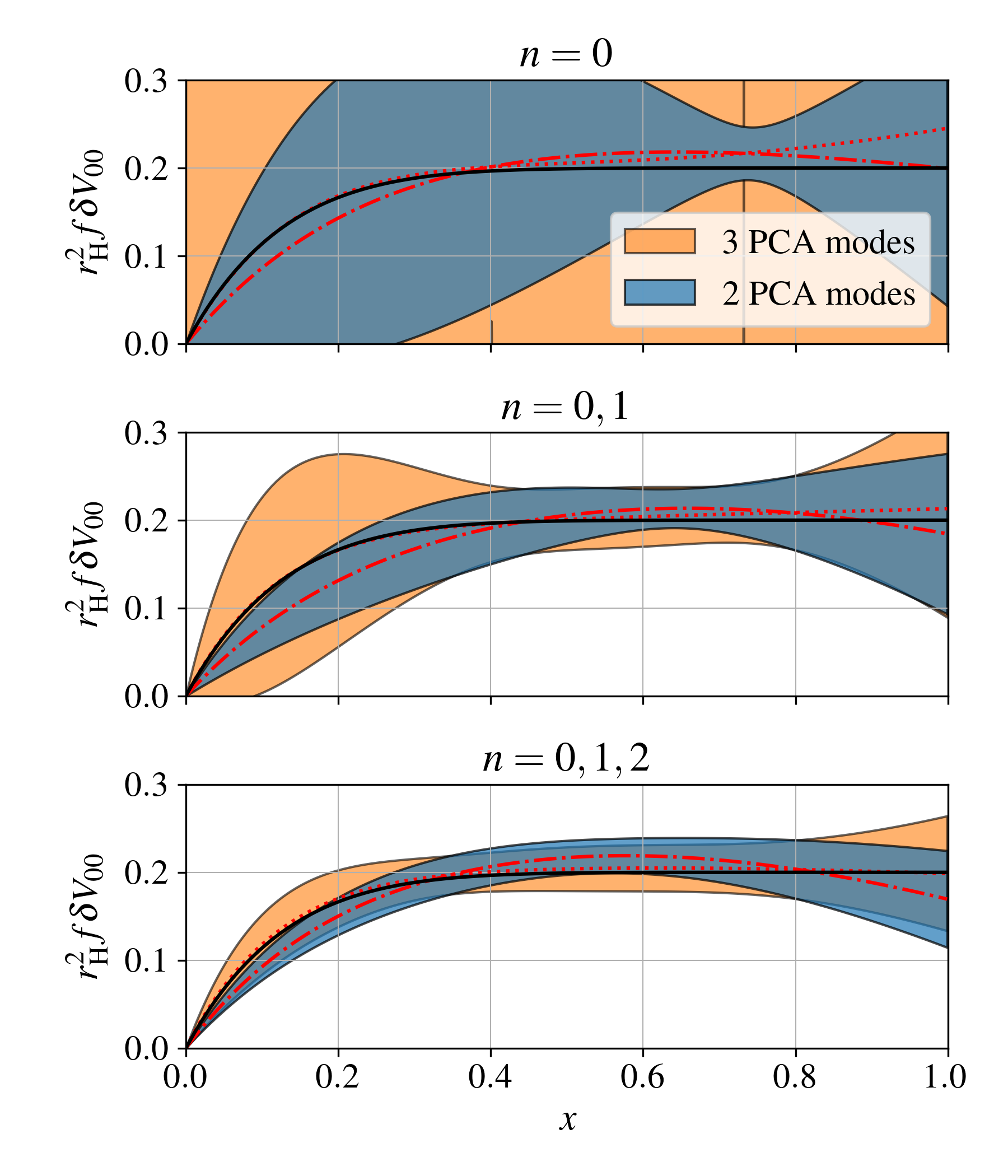}
\caption{Comparison of several PCA reconstructions (red lines) of an axial potential modification $\delta V_{00}$ (black solid). The associated PCA errors are indicated as colored areas. The different panels correspond to different numbers of  QNMs (top: $n=0$, middle: $n=0,1$, bottom: $n=0,1,2$). Red dashed lines correspond to the two most significant PCA components, red dotted lines to three.}
\label{PCAmodes}
\end{figure}

As a proxy, we use the same modification to the potential of section~\ref{app1}. In Fig.~\ref{PCAmodes} we show the PCA reconstruction of the problem when either the first two or three largest components $|b_k|/\sg_k$ are considered. Each panel of the figure replicates the reconstruction for a growing number of QNM modes used as observation, from one ($n=0$) to three ($n=0,1,2$).

The top panel, corresponding to only one observed QNM, shows that even if the reconstruction of the potential is rather close to the injection, the error bars are rather widespread. Moreover, one can notice that taking three PCA modes while having only two measured frequencies yields to a completely uninformative error.

In the mid and bottom panels, we show that the more frequencies we observe, the better the error bars become, as we already know from previous analysis. It is interesting to see that the reconstruction gets closer to the injection when more modes are considered, at a cost of having slightly larger error bars. This behaviour is expected, as the error is sensitive to the number of PCA components---{\it cfr.}~Eq.~\eqref{PCA_error}.

For the sake of clarity, we also analysed the inclusion of a fourth component in each, but it only marginally improves the results. This suggests that the information in the overtones, at least for this particular case, is not very significant in finding higher PCA modes. Our interpretation is that the information in the overtones is strongly ``correlated'' when looking into Fig.~\ref{qnm_art}, because the modifications of the QNM spectrum for different overtones is not perpendicular but systematically rather similar.


\end{document}